\documentclass[technote]{IEEEtran}
\usepackage{dblfloatfix} 
\newcommand{\tbl}[1]{Table~\ref{tbl:#1}}

\usepackage{amsmath}
\usepackage[table,xcdraw]{xcolor}
\usepackage{hyperref}

\usepackage{xcolor}

\usepackage{amssymb}

\ifCLASSOPTIONcompsoc
  \usepackage[nocompress]{cite}
\else
  \usepackage{cite}
\fi
\ifCLASSINFOpdf
\else
\fi

\usepackage{float}

\usepackage[framemethod=tikz]{mdframed}
\usetikzlibrary{shadows}
\newmdenv[
tikzsetting= {fill=gray!10},
linewidth=1pt,
roundcorner=2pt, 
shadow=false
]{myshadowbox}

\usepackage{graphicx}
\usepackage{subcaption}

\begin{document}
%
\title{Assessing Developer Beliefs: A Reply to ``Perceptions, Expectations, and Challenges in Defect Prediction''}

%
%
%
%

\author{Shrikanth N.C.,~\IEEEmembership{IEEE~Member} and
         ~Tim~Menzies,~\IEEEmembership{IEEE~Fellow}
\thanks{Shrikanth N.C. Tim   Menzies.
Computer
Science, NC State, USA.
Email: snaraya7@ncsu.edu and timm@ieee.org}
}
\markboth{IEEE TRANSACTIONS ON SOFTWARE ENGINEERING}
{Shell \MakeLowercase{\textit{et al.}}: Bare Demo of IEEEtran.cls for Computer Society Journals}
%



\newcommand{\bi}{\begin{itemize}}
\newcommand{\ei}{\end{itemize}}

\newcommand{\tion}[1]{\S\ref{tion:#1}}

\IEEEtitleabstractindextext{%
\begin{abstract} 
It can be insightful to extend qualitative studies with a secondary quantitative analysis
(where the former suggests insightful questions
that the latter can answer). 
Documenting developer beliefs should be the start, not the end, of Software Engineering research. Once prevalent beliefs are found,
they should be checked against real-world data.  
For example, this paper finds several
notable
discrepancies between 
empirical evidence and the
     developer beliefs    documented in Wan et al.'s recent TSE paper
     ``Perceptions,  expectations,  and  challenges  in  defectprediction''.     
By reporting these discrepancies
we can  stop developers (a)~wasting time on inconsequential matters or (b)~ignoring important effects. 

For the  future, we  would  encourage more   ``extension studies'' of prior qualitative results with quantitative empirical evidence.
\end{abstract}

\begin{IEEEkeywords}
defects, beliefs,  empirical software engineering
\end{IEEEkeywords}}

\maketitle

\IEEEdisplaynontitleabstractindextext

\IEEEpeerreviewmaketitle

\vspace{1cm}

\IEEEraisesectionheading{\section{Introduction}\label{sec:introduction}}
Just because software developers 
say they believe in ``X'',
that does not necessarily mean  that ``X'' is true.
J{\o}rgensen \& Gruschke~\cite{jorgensen09}
note that     software engineering   seldom uses lessons from past projects to improve their future reasoning (to the detriment  of new projects). 
Passos et al.~\cite{passos11} note that developers often assume that   lessons   learned from a few past projects are general to all  future projects~\cite{passos11}. 
Devanbu et al. record
 opinions about software development
  from 564 Microsoft software developers from around the world~\cite{prem16}. They comment 
  that programmer beliefs can (a)~vary with each project; and (b)~may  not necessarily
correspond with actual evidence in their current projects.
 
\begin{figure}
\scriptsize
    \centering
    \begin{subfigure}[b]{0.15\textwidth}
        \includegraphics[keepaspectratio,width=3cm]{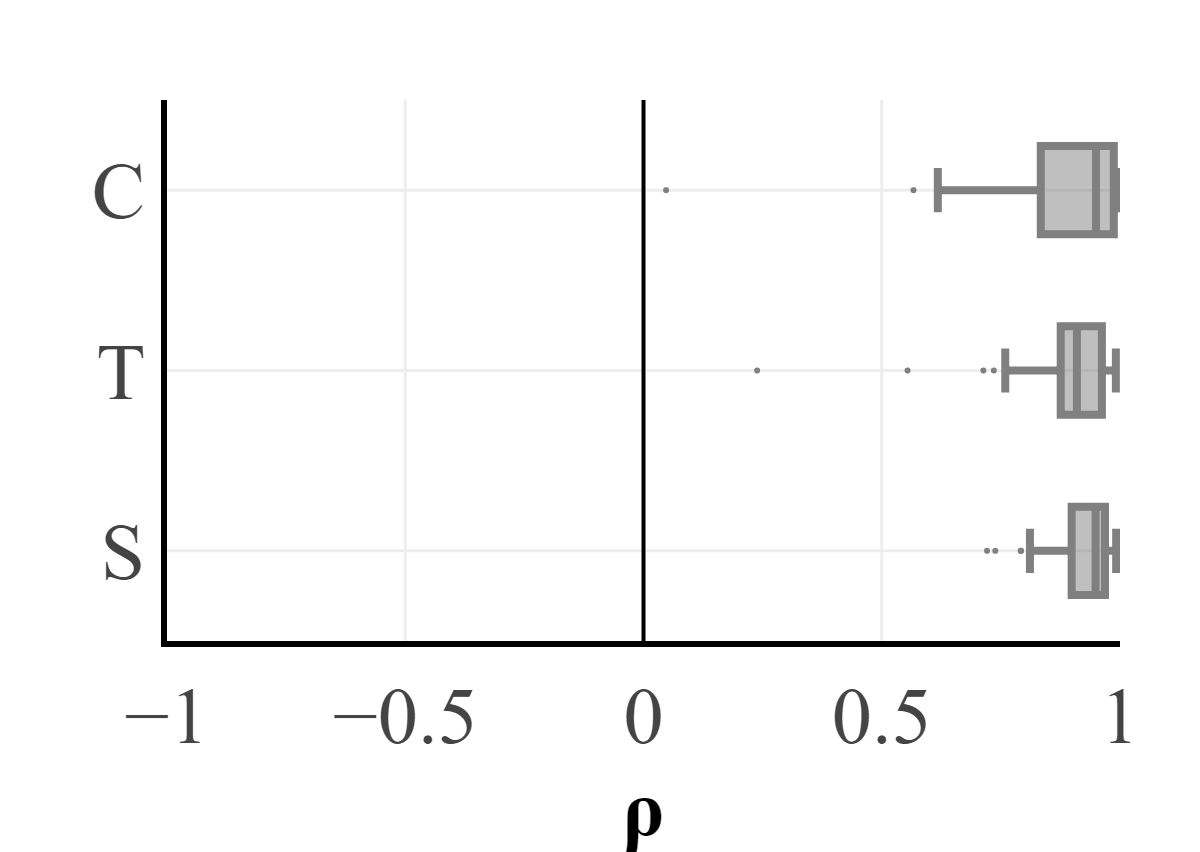}
        \caption{B6}
        \label{fig:b6}
    \end{subfigure}  ~ 
   \begin{subfigure}[b]{0.15\textwidth}
        \includegraphics[keepaspectratio, width=3cm]{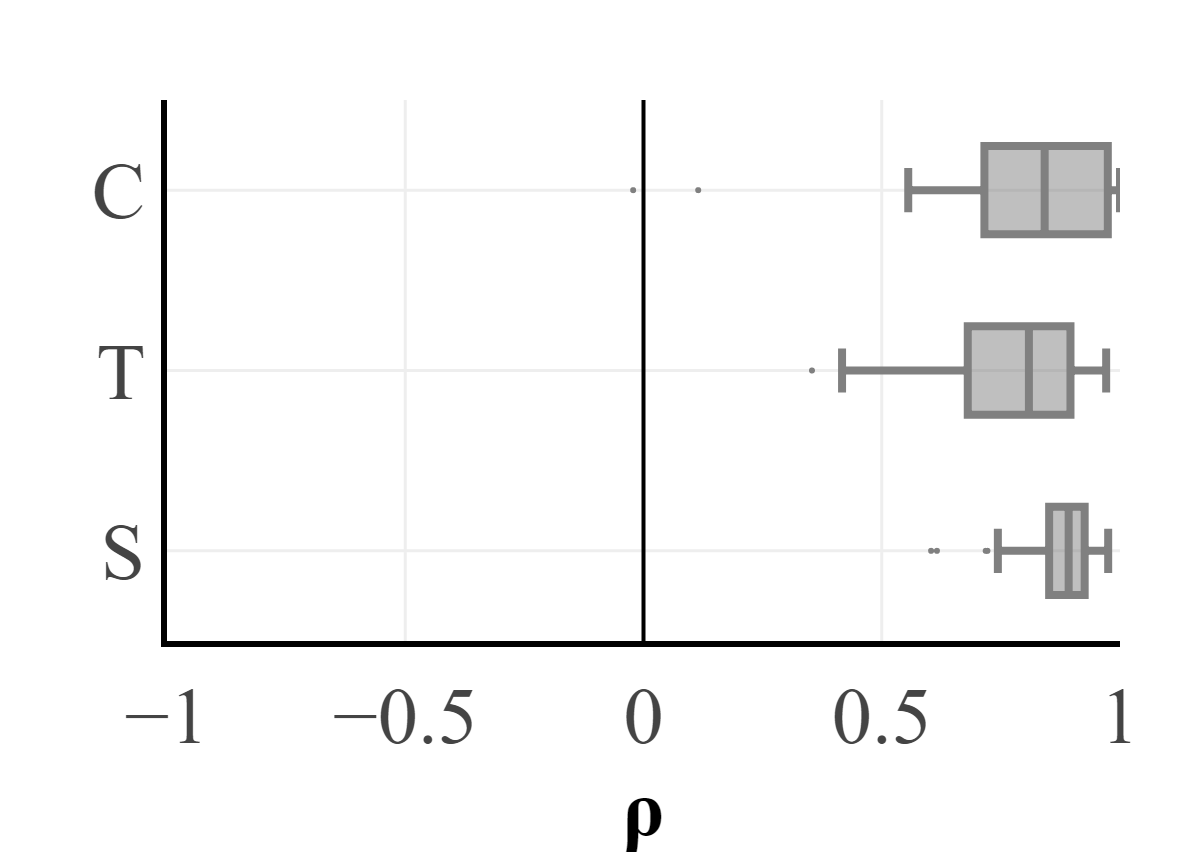}
        \caption{B1}
        \label{fig:b1}
    \end{subfigure}  ~ 
    \begin{subfigure}[b]{0.15\textwidth}
         \includegraphics[keepaspectratio, width=3cm]{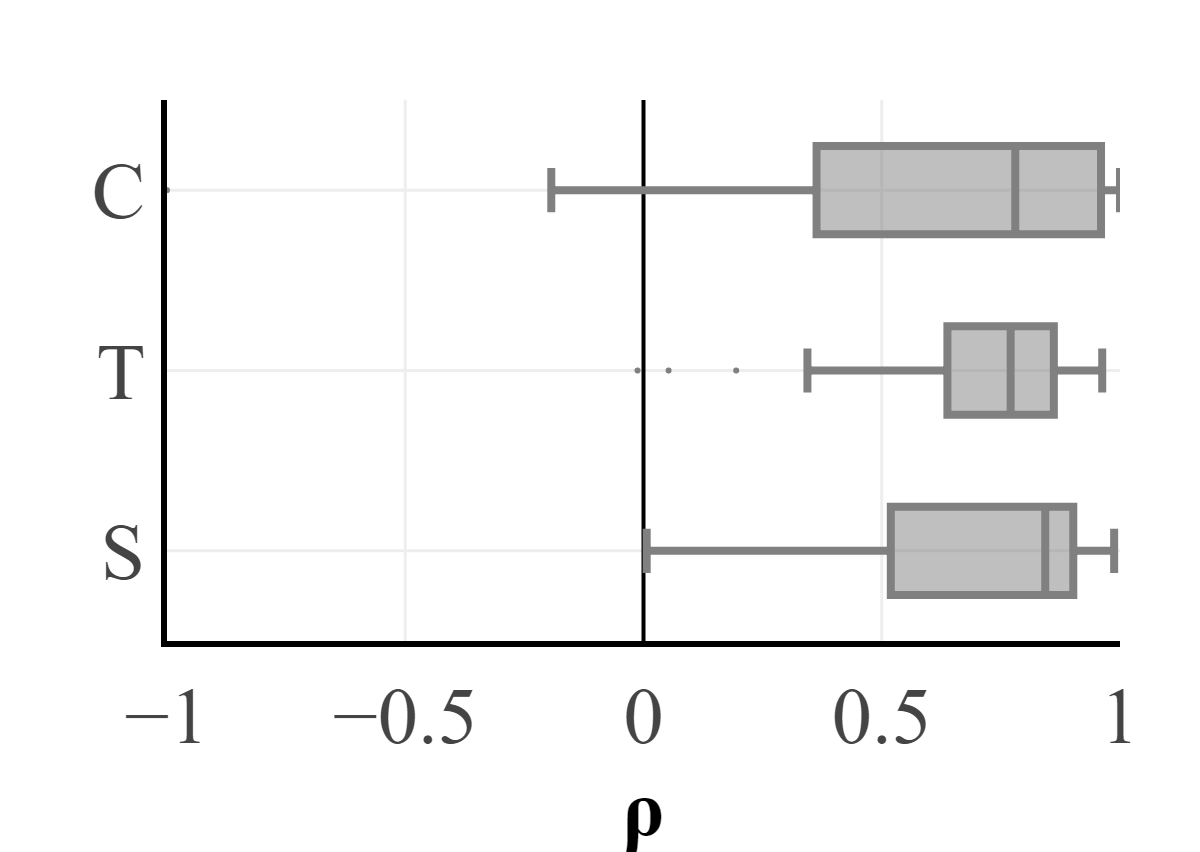}
        \caption{B4}
        \label{fig:b4}
    \end{subfigure}  ~ 
   \begin{subfigure}[b]{0.15\textwidth}
        \includegraphics[keepaspectratio, width=3cm]{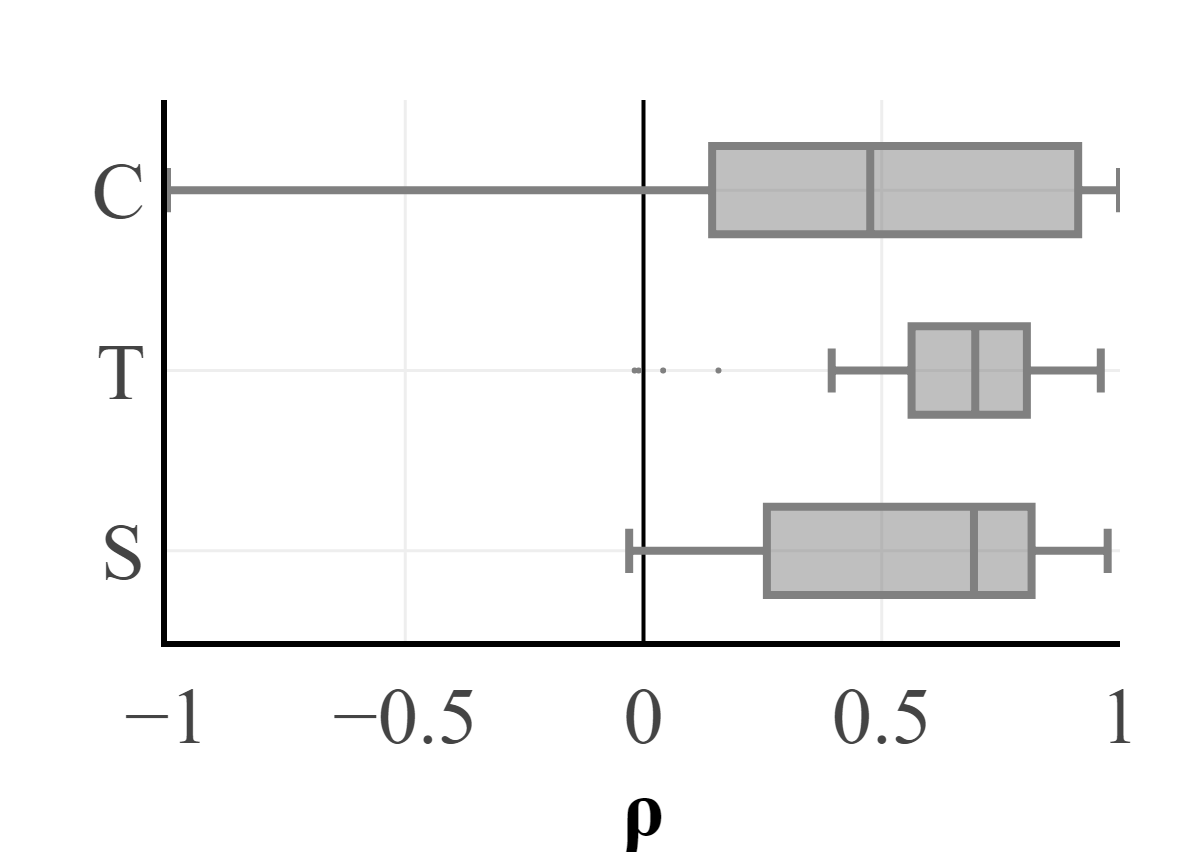}
        \caption{B2}
        \label{fig:b2}
    \end{subfigure}  ~ 
      \begin{subfigure}[b]{0.15\textwidth}
        \includegraphics[keepaspectratio, width=3cm]{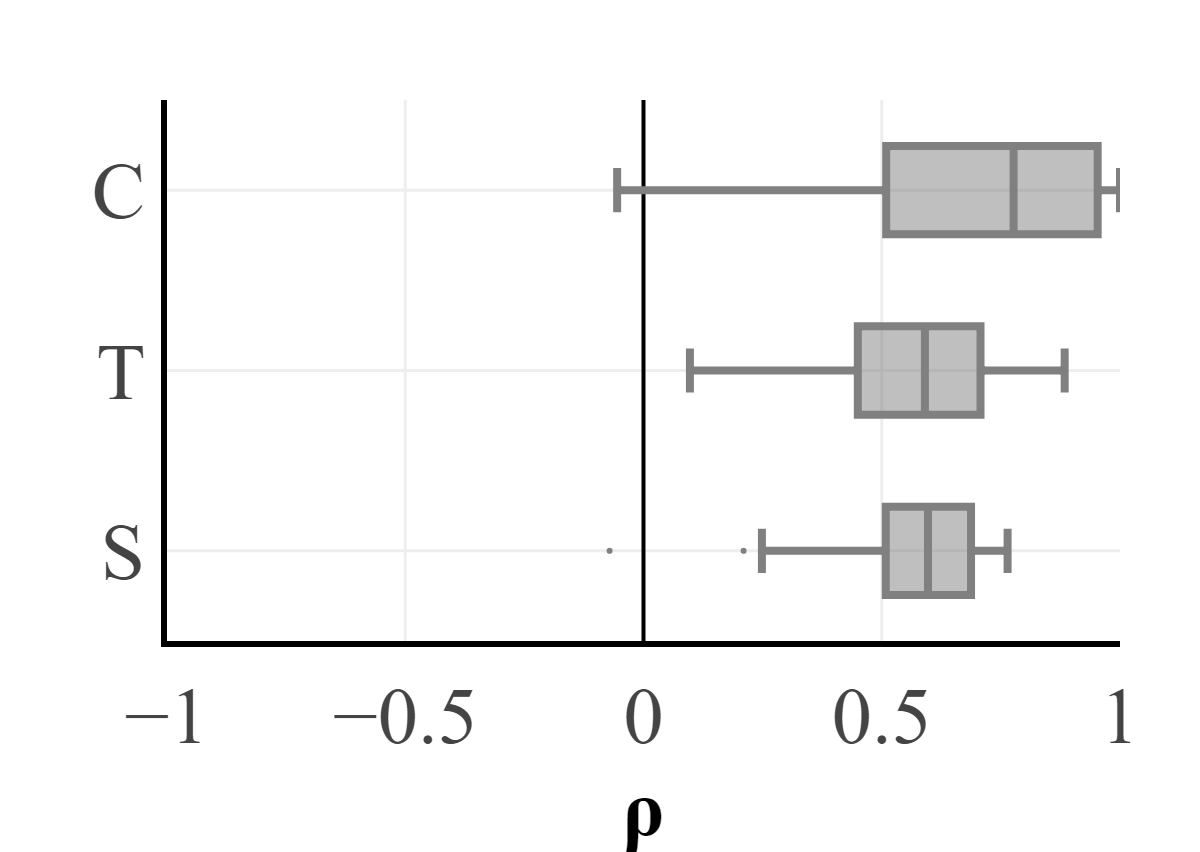}
        \caption{B8}
        \label{fig:b8}
    \end{subfigure}  ~ 
      \begin{subfigure}[b]{0.15\textwidth}
        \includegraphics[keepaspectratio, width=3cm]{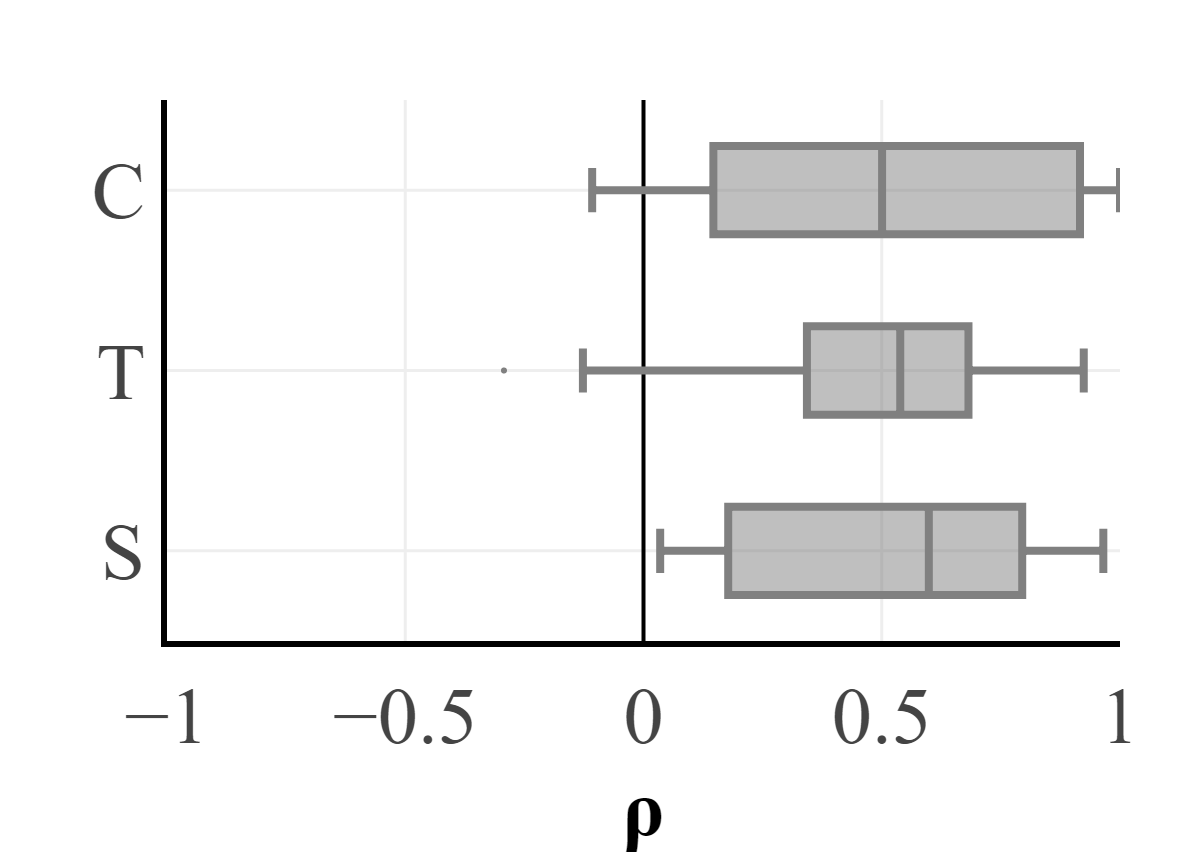}
        \caption{B7}
        \label{fig:b7}
    \end{subfigure}  ~ 
      \begin{subfigure}[b]{0.15\textwidth}
        \includegraphics[keepaspectratio, width=3cm]{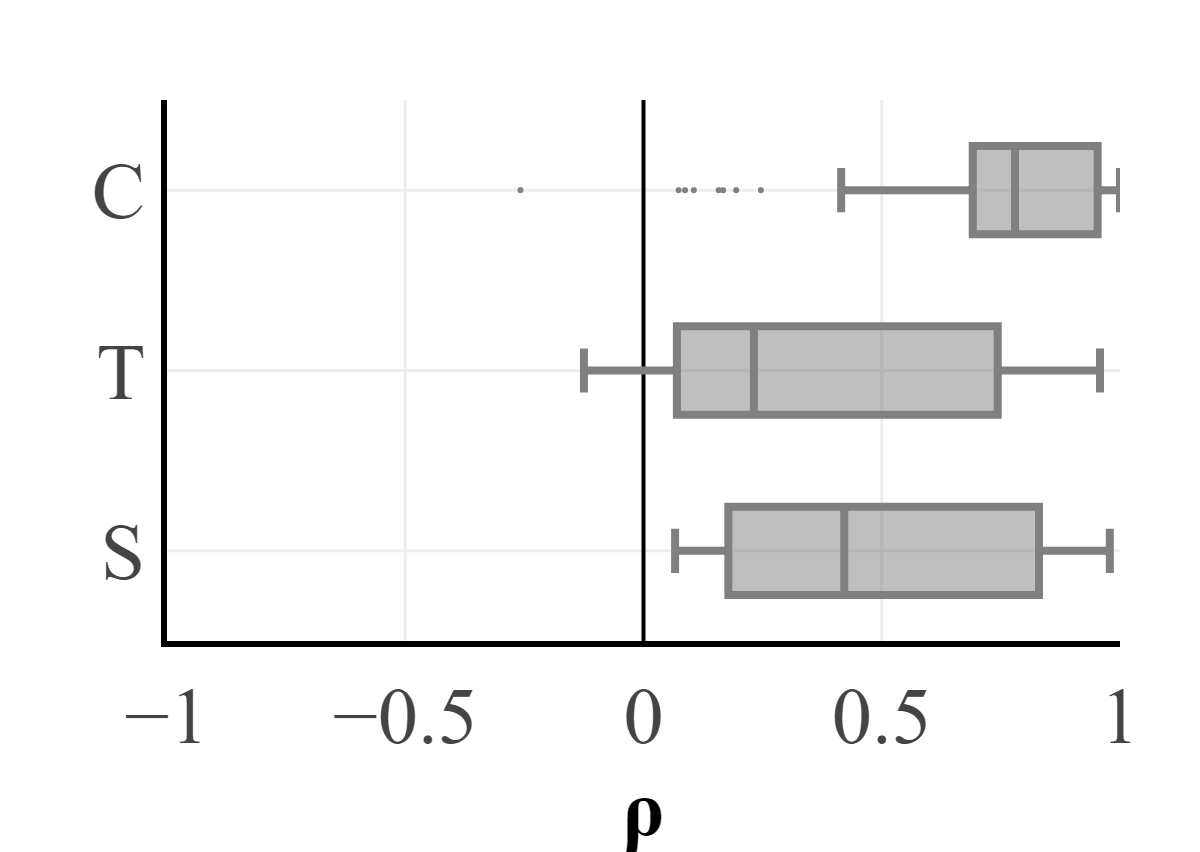}
        \caption{B5}
        \label{fig:b5}
    \end{subfigure}   ~       \begin{subfigure}[b]{0.15\textwidth}
        \includegraphics[keepaspectratio, width=3cm]{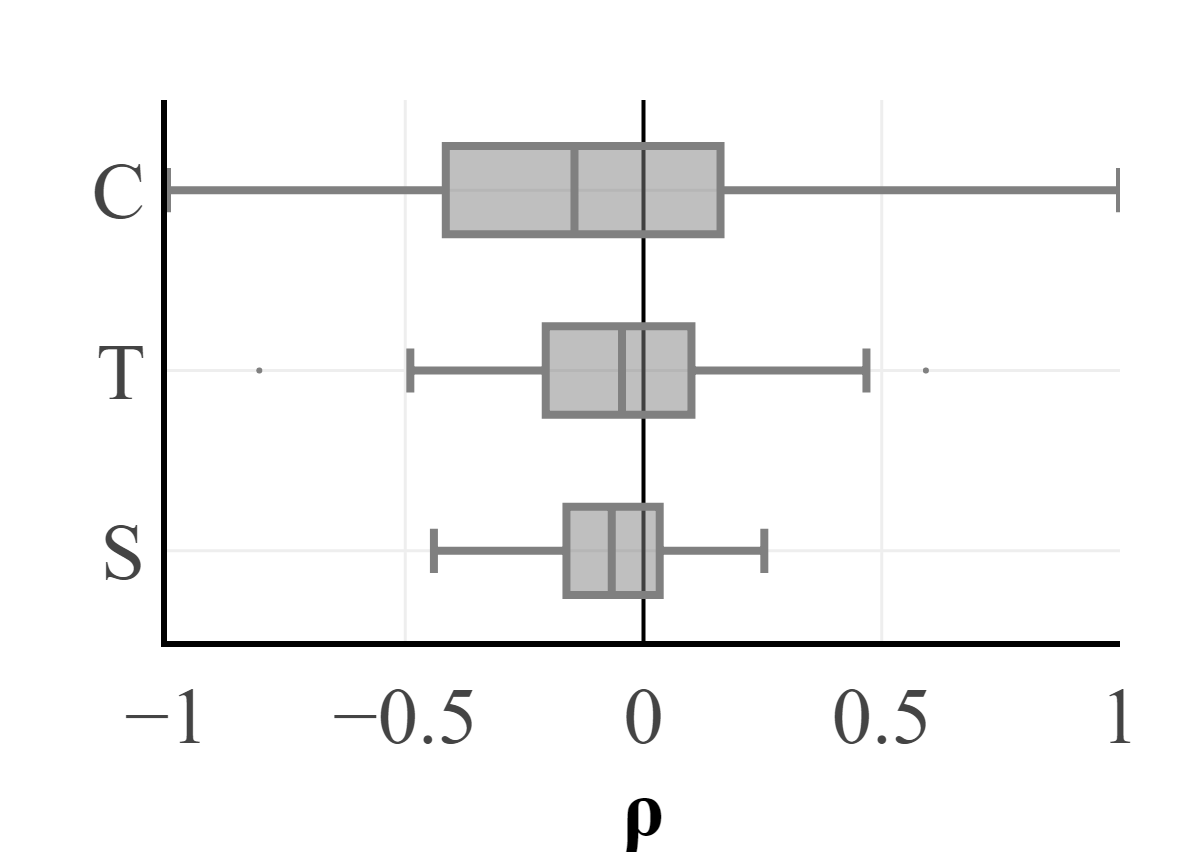}
        \caption{B3}
        \label{fig:b3}
    \end{subfigure}  
    \caption{ Correlation results for beliefs from \tbl{beliefs}. Sorted by median correlation.  Each sub-figure (a - h) shows 3 distributions computed using 2,670 Project Configuration files (\textbf{C}), 9,216 Test Case Files (\textbf{T})  and 27,806 Source Code Files (\textbf{S}) found in 46 projects.} 
\label{fig_boxplots}
\end{figure}

Accordingly,  we think it is important
to  evaluate   developer  beliefs reported 
in (e.g.)
Wan et al.'s
recent TSE paper
``Perceptions, expectations, and challenges in defect prediction''~\cite{cite_perceptions}.
That study collected 395 responses from practitioners
to document   developer beliefs about  willingness to adopt technologies, challenges, defect prediction metrics, etc.
Some of those beliefs about {\em defect prediction} can be tested empirically via  correlations to project data:
\bi
\item
In \tbl{beliefs},
the {\em \%agree} columns shows a {\em strength}
number that is {\em larger} when {\em more} developers
believe   something (in this table seven beliefs are based on prediction metrics, while B4 is based on prioritization strategies). 
\item
Figure~\ref{fig_boxplots}
shows  correlation between 
the beliefs in \tbl{beliefs} 
and data from 46 projects.

\ei
Note that the  beliefs  
do not correspond with the empirical evidence.
For example, 
  B6 has the highest correlation  in the data but it ranked bottom half in
\tbl{beliefs}.  Also, B3 has the lowest correlations in our data
 yet it is ranked in the top half of 
\tbl{beliefs}. Other issues are discussed later in \tion{results}.

\begin{table}[!t]

{\footnotesize
\begin{tabular}{|l@{~}|p{65mm}|@{~}r|}
\hline
\rowcolor{gray!30} 
\textbf{\#} & \multicolumn{1}{c|}{\textbf{Belief}}                                              & \multicolumn{1}{|l}{\textbf{ \% Agree}} \\ \hline
\textbf{B1}   & S14:\textit{Files changed by more developers are more buggy.}                                                             & \textbf{64}                                                     \\ \hline
\textbf{B2}   & \begin{tabular}[c]{@{}l@{}}S4:\textit{A file with more added lines is more bug-prone}\end{tabular} & \textbf{61}                                                     \\ \hline
\textbf{B3}   & S9:\textit{Recently created files tend to be buggy}                                                           & \textbf{52}                                                     \\ \hline

\textbf{B4}   & T7:\textit{A file with more Lines of Code (LOC)}                                                                       & \textbf{48}                                                     \\ \hline

\textbf{B5}   & S11:\textit{Files with more fixed bugs are more bug-prone}                                                        & \textbf{48}                                                     \\ \hline
\textbf{B6}   & S12:\textit{A file with more commits is more bug-prone}                                                               & \textbf{46}                                                     \\ \hline
\textbf{B7}   & S13:\textit{A file with more removed lines is more bug-prone}                                       & \textbf{35}                                                     \\ \hline
\textbf{B8}   & S15:\textit{Files with fewer lines contributed by their owners
(who contribute most changes) are more bug-prone}                                            & \textbf{30}                                                     \\ \hline
\end{tabular}
}
\caption{Developer beliefs, sorted by percent of developers
who endorsed that belief.  From Wan et al.~\cite{cite_perceptions}. 
}
  \label{tbl:beliefs}
\end{table}

\begin{table*}[!b]
\resizebox{\textwidth}{!}{%
\scriptsize
\begin{tabular}
{|l|l|r|l|l|r|l|l|r|}

\hline
\rowcolor{gray!30}
\textbf{github.com/}                    & \textbf{*Language} & \textbf{Commits} & \textbf{github.com/}                 & \textbf{*Language} & \textbf{Commits} & \textbf{github.com/}                     & \textbf{*Language} & \textbf{Commits} \\ \hline
activeadmin/activeadmin         & Ru CSS             & 6373             & karmi/retire                 & Ru Ht             & 996              & resque/resque                    & Ru Js             & 2487             \\ \hline
activemerchant/
active\_merchant & Ru Ht             & 4482             & lra/mackup                   & Python                      & 1914             & restsharp/RestSharp              & CSS Ht             & 1613             \\ \hline
aws/opsworks-cookbooks          & Ru Erb             & 2047             & lusis/chef-logstash          & Ru              & 790              & ros-simulation/gazebo\_ros\_pkgs & C++ Python                & 1398             \\ \hline
boto/boto3                      & Python Ht             & 2179             & Lusitanian/PHPoAuthLib       & PHP Ht            & 833              & Ru-grape/grape                 & Ru Ht             & 2170             \\ \hline
bundler/bundler                 & Ru Ht             & 11724            & mikel/mail                   & Ru              & 1901             & scribejava/scribejava            & Java          & 954              \\ \hline
cakePHP/phinx                   & PHP CSS             & 2304             & mperham/sidekiq              & Ru Js               & 3866             & Seldaek/monolog                  & PHP                     & 2119             \\ \hline
Codeception/Codeception         & PHP Ht            & 6625             & omniauth/omniauth            & Ru Ht             & 1385             & sferik/rails\_admin              & Js Ru               & 4485             \\ \hline
django-tastypie/django-tastypie & Python Ht             & 1335             & ooici/coi-services           & Python C                & 12982            & spotify/luigi                    & Python Js               & 3751             \\ \hline
doorkeeper-gem/doorkeeper       & Ru Erb             & 1798             & pennersr/django-allauth      & Python Ht             & 2058             & swanson/stringer                 & Ru              & 934              \\ \hline
drapergem/draper                & Ru Ht             & 1165             & pentaho/data-access          & Java Js             & 2575             & teamcapybara/capybara            & Ru              & 3756             \\ \hline
encode/django-rest-framework    & Python Ht             & 8229             & plataformatec/simple\_form   & Ru              & 2060             & thoughtbot/bourbon               & CSS Ht           & 1439             \\ \hline
errbit/errbit                   & Ru Ht             & 2813             & propelorm/Propel2            & PHP Ht            & 4343             & thoughtbot/paperclip             & Ru Ht             & 2026             \\ \hline
exercism/exercism.io            & Ru              & 6911             & puppetlabs/beaker            & Ru Shell               & 5127             & wlanslovenija/PiplMesh           & Python Ht             & 1660             \\ \hline
getpelican/pelican              & Ht Python             & 3092             & puppetlabs/puppetlabs-apache & Ru              & 3224             & xetorthio/jedis                  & Java Ht           & 2140             \\ \hline
irungentoo/toxcore              & C C++               & 3800             & reactjs/react-rails          & Js Ru               & 1080             & ZF-Commons/ZfcUser               & PHP SQL             & 887              \\ \hline
jordansissel/fpm                & Ru  Shell & 3885             & \multicolumn{6}{l}{    }                                                                                                                                    \\\cline{1-3}
\end{tabular}
}

\caption{46 Projects developed in various programming languages and of varying sizes (commits). We present the top 2 programming languages dominated by quantity in Language column (if applicable). \textbf{*Language  Ht = HTML , Js = JavaScript, Ru = Ruby and Erb = Embedded Ruby}. This data is available on-line at \href{https://github.com/ai-se/defect\_perceptions}{\textit{github.com/ai-se/defect\_perceptions}}.  }\label{tbl:projects}

\end{table*}

Before going any further, we  stress that while we doubt some of the {\em answers} offered
by Wan et al., we do not doubt the value of the {\em questions} they ask.
Software
analytics research needs to mature to the point where it can offer a set of $N$ conjectures
(that can be quickly tested) about what might reduce software project quality.
The analysis of Wan et al. is an important step towards that goal.

The other point to stress about 
Wan et al. is that it is an exemplary model
of how to do large scale qualitative Software Engineering (SE)
research. That said, that paper did not empirically assess the beliefs it documented. As shown by the following, such an assessment
can be an insightful quantitative extension to an initial qualitative study.
In the future, we would encourage more such dual qualitative+quantitative studies where the former suggests insightful questions
that the latter can explore. For other examples of this dual approach, see~\cite{chen2019replication,MenziesS03}.

The rest of this paper discusses how we tested the Wan et al. results and what was learned in the process.

\section{Method}

To test the beliefs in Table \ref{tbl:beliefs} we used
a   corpus from a recent ESEM'18 paper~\cite{cite_pareto}
(and that sample was drawn from projects labelled ``most suited(popular and active)'' in Github).
From this, we  randomly selected 50 projects in accordance
with development  language popularity~\cite{cite_top_prog} (so the projects under consideration
were developed in C, C\#, C++, Java, JavaScript, PHP, Python, Ruby, Shell, and HTML-CSS).

Initially, we targeted 50 projects since that was all we   could   reasonably present 
in a short TSE paper. After some data quirks, that resulted in the 46 projects of  \tbl{projects}. This sample contains
data   modified in the period 2005 to 2019 by  13,821 developers in
  592,094 file entries. These modifications were made to 
145,715   active branch commits (in all,
21,760,416 line insertions and 14,992,194 line deletions).

The kinds of files we found in each project were very varied. Lest that variation complicated the analysis, we divided the files into several, non-overlapping, categories:
\bi
\item
Source code files ending in .c, .cpp, .java etc;
\item 
Test cases files whose  file/path names include ``test'';
\item
Configuration files whose names end in .yml, .pom, etc.
\ei

Commits  were labelled ``bug fixing'' if they contained any derivatives of the following stemmed words like {\em bug, fix, issu, error, correct, proper, deprecat, broke, optimize, patch, solve, slow, obsolete, vulnerab, debug, perf, memory, minor, wart, better, complex, break, investigat, compile, defect, inconsist, crash, problem} or {\em resol} in their commit message. We found a minimum of 13\% and an average of 30\% bug(defect) fixing commits among the 46 projects we use in this study.

We say a file's {\em defect proneness}'  ``$D$'' is the number of times it was
changed(committed) for the purpose of fixing a defect. 

The correlation between $D$ and project attributes was checked using 
the Pearson's correlation
\mbox{$\rho = \frac{\text{cov}(X,Y)}{\sigma_x \sigma_y}$}
between  two samples $X,Y$ 
(with means $\overline{x}$
and $\overline{y}$), as estimated using $x_i \in X$ and $y_i \in Y$ via
\[\rho = \frac{{}\sum_{i=1}^{n} (x_i - \overline{x})(y_i - \overline{y})}{\sqrt{\sum_{i=1}^{n} (x_i - \overline{x})^2(y_i - \overline{y})^2}}\]
Our study then used advice from~\cite{cite_more_faults,cite_recently_created,cite_recently_age,cite_ownership}, as  follows:
\bi
\item For {\em B1:More developers},
we correlated  $D$ to the number of unique developers who made non-zero changes (line insertions or deletions) to a file.
\item
For {\em B2:Added lines} we  correlated  $D$ to  the number of lines added to a file during commit.
\item
For {\em B3:Recently Created},
the premise here  is that newly created files quickly   introduce new defects \cite{cite_recently_created}. To test this, we correlated creation time (larger value $\Rightarrow$ recent)  of a file's ``AGE''\cite{cite_recently_age} with the time interval between created time and the time of first defect fixing commit.
\item
For {\em B4:LOC} we correlated $D$ to a file's lines of code (additions + deletions throughout its commit history).
\item
For {\em B5:More Bug Fixes},
Similar to the approach in \cite{cite_more_faults},
we
split a file's commit history into two equal halves after sorted by time in ascending order. We then 
reported the correlation between the number of defect fixes in the first half with the number of defect fixes in the second half. 
\item
For {\em B6:More Commits}, we correlated $D$ to the number of commits for a file.
\item 
For {\em B7:Removed Lines},
the concern is that deleting too many lines in a file could make it defect prone. To test this,
we correlated $D$ and the number of lines deleted in a file during a commit.
\item
For {\em B8:Ownership}, we used Bird et al. \cite{cite_ownership}'s idea of
major and minor contribution. For each file,
we correlated   $D$ to
 \% of developers who wrote  $< 5 \%$ of lines. 
\ei 

\section{Results}\label{tion:results}

The first row of  Figure~\ref{fig_boxplots} shows that three
beliefs B6,B1,B4 are supported by the data (as witnessed 
by their high median correlations).
The B1 finding (that {\em more developers leads to more bugs})
is definitely consistent with the \tbl{beliefs} results.
However, this high support for B6 (that {\em more commits means more bugs})  is somewhat at odds with \tbl{beliefs} since, in that table,
developers endorse B6 less than half the time.
The clear message from B1 and B6 is that the more we have to tinker with code, the more bugs will be found.
It is a question for future work if such tinkering is {\em necessitated} by the presence of bugs
or whether or not that tinkering {\em causes} the bugs.

As to B4 (that {\em large files have more bugs}),
this result is somewhat equivocal.
The  larger range of $\rho$ in the results mean that there will
be several projects where larger files do not have no more bugs.
So we say that B4 has somewhat weaker support in the data 
 than B1 and B6.
 
At the other end of the scale, our results   offer   little support for
B3
(that {\em recently changed files tend to be buggy});
B5 (that {\em files with more fixed bugs are more bug-prone});
and  B7 (that {\em files with more lines removed is more bug prone)}.
All these beliefs show poor correlations to empirical data,
especially B3 (whose correlations are often found near $\rho = 0$).
The B5 and  B7 results are consistent with Wan et al.'s developer beliefs
(who did not rate this belief highly in \tbl{beliefs}). However, the B3 results are a
 discrepancy
since developers in \tbl{beliefs} agreed with this belief at least half the time.

As to other beliefs, 
these exhibit many correlations below accepted thresholds
for ``strong'' correlation~\cite{asuero2006correlation}.
Hence we cannot support
B2 (that {\em a file with more added lines is more bug-prone})
or B8 (that {\em prevalence of changes by file owners
  decreases bugs}).
The B8 results are consistent with \tbl{beliefs}.
However,
the B2 results are a discrepancy since it is ranked higher in \tbl{beliefs}.

\section{Threats to validity}

Any data mining project is prone to {\em sample bias} where the conclusions are  distorted by the data used to make the conclusions.
To mitigate that problem, we have explored a large sample of projects.
Also, all our data is on-line so  that other researchers can check for data distortions (see
\href{https://github.com/ai-se/defect\_perceptions}{\textit{github.com/ai-se/defect\_perceptions}}).

Another kind of {\em sampling bias} is ``cherry picking'' which beliefs to explore and which to ignore. This paper only explores 8 of the 33 beliefs documented by Wan et al. We found that some of the modeling    decisions about how to map   data into \tbl{beliefs}   required extensive, possibly even arcane, explanations.
Accordingly,   we elected  to use just the \tbl{beliefs} beliefs since these  could be easily  mapped 
into   data via  simple correlation.  We assert that we finalized the list of beliefs in \tbl{beliefs} {\em before}
checking for correlations.
That is, we did not maliciously ``cherry pick'' just a set of beliefs that have some discrepancies with empirical data.

This work required several modeling decisions in order to map the available data into \tbl{beliefs}.
For
example:
\bi
\item
For belief {\em B3:Recently Created},
it is possible that a file may be involved in fixing defects multiple times in its lifetime, especially older files. In such cases we consider the interval between the   two most recent defect fixes. 
\item
We do not analyze static files (such as image, text etc) since we think it safe to assume that these are  not used in fixing software bugs.
\item
A very small number of files ($\le 5$\%), had no reported defects (especially for the configuration files).
These files were grouped into the little set of Figure~\ref{fig_boxplots} since such files   offer  no support for any belief.
\item To make our conclusions, we had to decide what was a high correlation. Using advice from the  literature~\cite{asuero2006correlation}, we used $\rho > 0.7$.
We acknowledge 
that the decision is debatable. 
\item Etc.
\ei
These decisions introduce a threat to validity  (i.e. if those modeling decisions were wrong, then so are our conclusions).
In order to mitigate that problem,  where possible, we took advice from recent research papers~\cite{cite_more_faults,cite_recently_created,cite_recently_age,cite_ownership}.

Finally, Wan et al.   did not explore language-specific stratifications of the data. Since we are comparing our results to theirs, we also 
did not check the effect of programming languages
on the  beliefs of \tbl{beliefs}
(but that might be an insightful extension for future work).

\section{Conclusion}
At the start of a software analytics project,
it is important to focus software analytics
on questions of interest to the client.
Therefore, it is  very
 important to document developer beliefs, as done by Wan et al. 

 That being said, once project data becomes available, it is just
 as important to update the focus in accordance with the observed
 effects.  
 
 Going forward, this work prompts us to
 explore better  tool support for {\em belief
 revision} in software analytics. There must
  be some way to politely, yet convincingly, encourage
 developers to update their beliefs when evidence demands it.
 In this way, we can stop developers  wasting time on inconsequential matters (e.g.B2,B3),
 or ignoring important effects (e.g. B6).

\bibliographystyle{abbrv}
\bibliography{references}




%





\end{document}